\documentclass[aps,prd,noshowpacs,eqsecnum,nofootinbib,twocolumn]{revtex4}

\usepackage{longtable}

\usepackage{latexsym}

\usepackage{amsmath}

\usepackage{array}

\usepackage{color}

\begin{document}


\title{Five Texture Zeros and CP violation for the SM Quark Mass Matrices}

\author{William A. Ponce}

\affiliation{Instituto de F\'isica, Universidad de Antioquia, A. A. 1226, Medell\'in, Colombia}


\author{Richard H. Benavides}

\affiliation{Instituto Tecnol\'ogico Metropolitano, Facultad de Ciencias, Medell\'in, Colombia}

\author{John D. G\'omez}

\affiliation{Instituto de F\'isica, Universidad de Antioquia, A. A. 1226, Medell\'in, Colombia}

\date{\today}

\begin{abstract}

The phenomenology of the five independent sets of $3\times 3$ quark mass matrices with five texture zeros is 

carried through in full detail, including predictions for the CP violation asymmetries. Our study is done without 

any approximation, first analytical and then numerical.

\end{abstract}



\maketitle



\section{Introduction}

Although the gauge boson sector of the Standard Model (SM) with the $SU(3)_c\otimes SU(2)_L\otimes U(1)_Y$  local symmetry 

has been very successful~\cite{SM}, its Yukawa sector is still poorly understood. Questions related with this sector as 

for example the total number of families in nature, the hierarchy of the charged fermion mass spectrum, the smallness of

neutrino masses, the quark mixing angles, the neutrino oscillations and the origin of the CP violation, remain until today

as open problems in theoretical particle physics.

The mechanisms for fermion mass generation and flavor mixing can be classified into four different categories: (i) radiative mechanisms~\cite{wein} (which includes the so-called Froggat and Nielsen mechanism~\cite{Fro}); (ii) texture zeros in the mass matrices which may predict self-consistent and experimentally favored relations between fermion masses and flavor mixing parameters~\cite{Fri,Ram}; (iii) family symmetries, discrete~\cite{Dis} and continuous, global and local gauge symmetries~\cite{Con}, and (iv) see-saw mechanisms for electrically neutral~\cite{seneu} and charged particles~\cite{sesw}, related to a natural interpretation of the smallness of some fermion masses.

In the SM and after the local gauge symmetry has been spontaneously broken, the quark mass terms are given by

\begin{equation}\label{lama}
-{\cal L}_M=\bar U_{0L}M_u U_{0R}+\bar D_{0L}M_dD_{0R}+h.c,
\end{equation}

where $\bar U_{0L}=(\bar u_0,\bar c_0,\bar t_0)_L,\;\; \bar D_{0L}=(\bar d_0,\bar s_0,\bar b_0)_L,\;\;
U_{0R}=(u_0,c_0,t_0)^T_R,\;\;D_{0R}=(d_0,s_0,b_0)^T_R,$ (the upper $T$ stands for transpose, and the down zero stands for weak basis states). The matrices  $M_u$ and $M_d$ in (\ref{lama}) are in general $3\times 3$ complex mass matrices. In the most general case they contain  36 free parameters. In the context of the SM, such a large number of parameters can be drastically cut by making use of the polar theorem of matrix algebra, by which, one can always decompose a general matrix as the product of an hermitian times a unitary matrix. Since in the context of the SM the unitary matrix can be absorbed in a redefinition of the right handed quark components, this immediately brings down the number of free parameters from 36 to 18 (the other eighteen parameters can be hidden in the right-handed quark components in the context of the SM and some of its extensions, but not in its left-right symmetric extensions).

So, as far as the SM is concerned, we may treat without loss of generality $M_u$ and $M_d$ as two hermitian quark mass matrices, with 18 real parameters in total, out of which six are phases. Since five of those phases can be absorbed in a redefinition of the quark fields~\cite{koma}, the total number of free parameters we may play with in $M_u$ and $M_d$ are 12 real parameters and one phase; this last one used to explain the CP violation phenomena.

But in the context of the SM it is always possible to implement the so called weak basis (WB) transformation, which leaves the two $3\times 3$ quark mass matrices Hermitian, and do not alter the physics implicit in the weak currents. Such a WB transformation is a unitary transformation acting simultaneously in the up and down quark mass matrices~\cite{Bra1}. That is

\begin{equation}\label{2aa}
\begin{split}
M_u&\longrightarrow M_u^R=U M_u U^\dag,\\
M_d&\longrightarrow M_d^R=U M_d U^\dag,
\end{split}
\end{equation}

where $U$ is an arbitrary unitary matrix. We say then that the two representations $(M_{u},M_{d})$ and $(M_{u}^R,M_{d}^R)$ are equivalent in the sense that they are related to the same $V_{CKM}$ mixing matrix. This kind of transformation plays an important role in the study of the so-called flavor problem.

In the last paper of Ref.~\cite{Bra1} it was shown that, related to the mass hierarchy $m_u<m_c<m_t$ and $m_d<m_s<m_b$, it is always possible to perform a weak basis transformation in the hermitian quark mass matrices such that

\begin{equation}\label{owbt}
\begin{split}
(M_u^R)_{11}=(M_d^R)_{11}=(M_u^R)_{13}=(M_u^R)_{31}=0;\\
{\rm or\;\; equivalently};\\
(M_u^R)_{11}=(M_d^R)_{11}=(M_d^R)_{13}=(M_d^R)_{31}=0.
\end{split}
\end{equation}

According to this, it is always possible to have hermitian quark mass matrices with three texture zeros which do not have any physical implication. With three texture zeros the number of free parameters in $M_u^R$ and $M_d^R$ reduces from twelve to nine real plus one phase, just enough to fit the measured values for the six quark masses, the three mixing angles, and the CP violation phenomena. Any extra texture zero can only be a physical assumption and should imply a relationship between the quark masses and the parameters of the mixing matrix. But the maximum number of such texture zeros consistent with the absence of a zero mass eigenvalue and a non degenerate spectrum is just six, with three in the up and three in the down quark sectors respectively.

In what follows we are going to present analytic and numeric results for the set of SM quark mass matrices containing five texture zeros, taking special care to accommodate the latest experimental data available~\cite{pdg}, including the CP violation phenomena.

This paper is organized as follows: in Sec.~\ref{sec:sec2} some features of the SM mixing matrix are presented, in Sec.~\ref{sec:sec3} we introduce the five independent sets of $3\times 3$ quark mass matrices with five texture zeros, in Sec.~\ref{sec:sec4} we present our analytic and numerical analysis. A discussion of the results is done in Sec.~\ref{sec:sec5} and the conclusions are presented in Sec.~\ref{sec:sec6}. Two appendices are written at the end, the first one presenting the experimental measured values used in the main text, and a second one is a pure mathematical appendix, dealing with the analytic diagonalization of real $3\times 3$ orthogonal matrices.


\section{\label{sec:sec2}The SM Mixing Matrix}

In the SM and for the six flavor case, the Baryon charged weak current is given by

\begin{equation}\label{weakcurr}
J_{\mu L}^-=\bar U_{0L}\gamma_\mu D_{0L}=\bar U_L\gamma_\mu V_{CKM}D_L,
\end{equation}

where  $V_{CKM}=U_u U_d^\dag$ is the Cabibbo-Kobayashi-Maskawa (CKM) mixing matrix, with $U_u$ and $U_d$ the unitary matrices which diagonalize the Hermitian $M_uM_u^\dagger$ and $M_dM_d^\dagger$ square mass matrices respectively, and $\bar U_{L}=(\bar u,\bar c,\bar t)_L$ and $D_{L}=(d,s,b)^T_L$ stand for the quark field mass eigenstates.

$V_{CKM}$ is a $3\times 3$ unitary matrix, its form is not unique, but the permutation freedom between the three generations can be  removed by ordering the families such that $(u_1,u_2,u_3)\rightarrow(u,c,t)$ and $(d_1,d_2,d_3)\rightarrow(d,s,b)$. The complex elements of $V_{CKM}$ are thus commonly written as

\begin{equation}\label{vdef}
 V_{CKM}=\left(\begin{array}{ccc}
V_{ud} & V_{us} & V_{ub} \\
V_{cd} & V_{cs} & V_{cb} \\ 
V_{td} & V_{ts} & V_{tb} 
               \end{array}\right).
\end{equation}

The unitary of the CKM mixing matrix leads to relations among the rows and columns of $V_{CKM}$, in particular we have for the columns:

\begin{subequations}\label{uttri}

\begin{align}\label{ts}
V_{ud}V_{us}^*+V_{cd}V_{cs}^*+V_{td}V_{ts}^*&=0, \\ \label{tc} 
V_{us}V_{ub}^*+V_{cs}V_{cb}^*+V_{ts}V_{tb}^*&=0, \\ \label{tb}
V_{ud}V_{ub}^*+V_{cd}V_{cb}^*+V_{td}V_{tb}^*&=0. \\ \nonumber
\end{align}

\end{subequations}

Each of these three relations requires the sum of three complex quantities to vanish and so can be geometrical represented in the complex plane as a triangle. These are the unitary triangles~\cite{nir}, though the term ``unitary triangle is usually reserved for the relation (\ref{tb}) only.

The three angles of the unitary triangle  represented by (\ref{tb}), which are physical quantities and can be independently measured by CP asymmetries in B decays. are defined as follows~\cite{nir}:

\begin{subequations}\label{cpangl}

\begin{align}\label{alpha}
\alpha&\equiv{\rm arg}\left[-\frac{V_{td}V^*_{tb}}{V_{ud}V_{ub}^*}\right],\\ \label{beta}
\beta&\equiv{\rm arg}\left[-\frac{V_{cd}V^*_{cb}}{V_{td}V_{tb}^*}\right],\\ \label{gamma}
\gamma&\equiv{\rm arg}\left[-\frac{V_{ud}V^*_{ub}}{V_{cd}V_{cb}^*}\right].\\ \nonumber
\end{align}
\end{subequations}

The experimental findings at the B factories, fitted to close the triangle, are~\cite{ckmf, babar}

\begin{equation}\label{utex}
(\alpha,\;\beta,\;\gamma)_{exp}^{fit}=(95.9^{+2.2}_{-5.6},\;\;\;21.8\pm 2.8,\;\;\;67.2^{+4.4}_{-4.6}),
\end{equation}

with an accuracy in the measurement of $\sin 2\beta$ no less than 20\%~\cite{sony}.

The Cabbibo-Kobayashi-Maskawa $V_{CKM}$ matrix, can be parameterized by three mixing angles $\theta_{ij},\;\; i<j,\;\; i,j=1,2,3,$ the angles between the $i^{th}$ and $j^{th}$ families, and only one CP violating phase~\cite{koma}. Of the many possible parameterizations, the standard choice is~\cite{pdg,smpar}

\begin{widetext}

\begin{equation}\label{vdspar}
 V_{CKM}=\left(\begin{array}{ccc}
c_{12}c_{13} & s_{12}c_{13} & s_{13}e^{-i\delta} \\
-s_{12}c_{23}-c_{12}s_{23} s_{13}e^{i\delta}&  c_{12}c_{23}-s_{12}s_{23}s_{13}e^{i\delta} & s_{23}c_{13} \\ 
 s_{12}s_{23}-c_{12}c_{23}s_{13}e^{i\delta}&  -c_{12}s_{23}-s_{12}c_{23}s_{13}e^{i\delta}& c_{23}c_{13} 
               \end{array}\right),
\end{equation}

\end{widetext}

but the most important fact related with this matrix is that most of its entries have been measured with high accuracy, with the following  

bounds~\cite{pdg,ckmf}:

\begin{widetext}

\begin{equation}\label{maexp}
 V^{(exp)}=
\left(\begin{array}{ccc}
0.970\leq |V_{ud}|\leq 0.976 & 0.222\leq |V_{us}|\leq 0.226 & 0.003\leq |V_{ub}|\leq0.004\\
0.217\leq |V_{cd}|\leq 0.237 & 0.960\leq |V_{cs}|\leq 0.990   & 0.039 \leq |V_{cb}|\leq 0.041\\
0.008\leq |V_{td}|\leq 0.009 & 0.038\leq |V_{ts}|\leq 0.042 & 0.999\leq |V_{tb}| < 1.000
\end{array}\right),
\end{equation}
\end{widetext}

where the experimental numbers quoted above at 95\% C.L. are beyond the pure experimental bounds (which call for example for a $|V_{tb}|>0.79$) because they are restricted to fit the unitary conditions of $V_{CKM}$. The numbers quoted in (\ref{maexp}) are the most convenient for our purpose, due to the fact that we are going to confront these numbers with quark mass matrices which must fit the SM constraints.


\section{\label{sec:sec3}Five texture zeros}

As mentioned before, the maximum number of texture zeros in the quark mass matrices of the SM, consistent with a non degenerate spectrum and with the absence of a zero mass eigenvalue, are three in the up and three in the down quark sectors respectively. With this in mind, Harald Fritzsch proposed some time ago~\cite{Fri} the existence of parallel three texture zeros structures for the $3\times 3$ quark mass matrices $M_u$ and $M_d$ of the SM, such that $(M_q)_{11}=(M_q)_{22}=(M_q)_{13}=(M_q)_{31}=0,\;\; q=u,d$. This original Fritzsch ``anz\"atze" is named today in the literature as the ``parallel nearest neighbor interaction form''. With six texture zeros there are left just six real parameters plus one phase to describe the six quark masses, the three mixing angles and the CP violation phase, and so, the three mixing angles of the CKM mixing matrix can be expressed as functions of the quark masses and of the CP violation phase.

As it has been shown in several places, the parallel and non-parallel six texture zeros does not work properly and it has been ruled out by analytic~\cite{ripo} and numeric~\cite{numal} studies (only a charm quark mass half of its measured value can rescue this "anz\"atze").

Later, Ramond, Roberts and Ross (RRR) gave up the parallelism between the structures of $M_u$ and $M_d$ and found that there exist five phenomenologically allowed patterns of Hermitian quark mass matrices, each one of them with five texture zeros~\cite{Ram}, as listed in the table.

\vspace{1cm}

\begin{center}

\begin{tabular}{c|c|c}\hline\hline\label{T1}

Pattern & $M_u^\#$ & $M_d^\#$ \\ \hline

I  &  $\left(\begin{array}{ccc}

0     & a_u    & 0    \\

a_u^* & c_u    & b_u  \\

0     & b_u^*  & d_u

\end{array}\right)$

 & $\left(\begin{array}{ccc}

0     & a_d   & 0    \\

a_d^* & c_d   & 0  \\

0     & 0     & d_d

\end{array}\right)$ \\

II   & $\left(\begin{array}{ccc}

0     & a_u   & 0    \\

a_u^* & c_u     & 0  \\

0     & 0 & d_u

\end{array}\right)$

 & $\left(\begin{array}{ccc}

0     & a_d   & 0    \\

a_d^* & c_d   & b_d  \\

0     & b_d^* & d_d

\end{array}\right)$

\\

III  & $\left(\begin{array}{ccc}

0     & a_u   & 0    \\

a_u^* & 0     & b_u  \\

0     & b_u^* & d_u

\end{array}\right)$

 & $\left(\begin{array}{ccc}

0     & a_d   & 0    \\

a_d^* & c_d   & b_d  \\

0     & b_d^* & d_d

\end{array}\right)$\\

IV   &  $\left(\begin{array}{ccc}

0     &  0    &  a_u   \\

0     & c_u   & b_u  \\

a_u^* & b_u^* & d_u

\end{array}\right)$

 & $\left(\begin{array}{ccc}

0     & a_d   & 0    \\

a_d^* & c_d     & 0    \\

0     & 0     & d_d

\end{array}\right)$ \\

V & $\left(\begin{array}{ccc}

0     &  0   & a_u     \\

0     & c_u  & 0       \\

a_u^* & 0    & d_u

\end{array}\right)$

 & $\left(\begin{array}{ccc}

0     & a_d   & 0    \\

a_d^* & c_d   & b_d  \\

0     & b_d^* & d_d

\end{array}\right)$ \\  \hline\hline

\end{tabular}

Five RRR patterns of Hermitian quark mass matrices.

\end{center}

\vspace{0.9cm}

A few remarks are in order:\\

i) Patterns I, II, IV and V have 3 different phases, so 2 are unphysical. Pattern III has four different phases, so 3 are unphysical.\\

ii) Each pattern has 7 real parameters which means two physical predictions. That is, for each pattern 2 of the 3 mixing angles can be written as a function of the six quark masses and of the CP violation physical phase.\\

iii) The five different RRR patterns commit at least with one (some with both) of the weak basis texture arrangements in Eqs.~(\ref{owbt}).

In what follows we are going to study in detail these independent RRR patterns, paying special attention for each case to the prediction for the Cabbibo angle and to the CP violation phenomena.


\section{\label{sec:sec4}Analytic and numeric analysis}

\subsubsection{Pattern I}

This pattern was studied previously in Ref.~\cite{ripo}. The complex Hermitian quark mass matrices for this pattern can be written as:

\begin{equation}\label{1up}
M_u^{wb}=\left(\begin{array}{ccc} 
0     & |a_u|e^{i\alpha_u^w}    & 0    \\ 
|a_u|e^{-i\alpha_u^w}  & c_u    & |b_u|e^{i\beta_u^w}   \\
0     & |b_u|e^{-i\beta_u^w}   &  d_u 
\end{array}\right),
\end{equation}

\begin{equation}\label{1dw}
M_d^{bf}=\left(\begin{array}{ccc} 
0     & |a_d|e^{i\alpha_d^b} & 0    \\ 
|a_d|e^{-i\alpha_d^b}  & c_d   & 0  \\
0     & 0     & d_d
\end{array}\right).
\end{equation}

As can be seen, for this pattern the mixing angles $\theta_{13}$ and $\theta_{23}$ between the third family and the first two

 ones, come only from the up quark sector. The complex phases are removed by going to a prime basis using the following unitary transformations:

\begin{equation}\label{upph}
\left(\begin{array}{c} u_0^\prime \\c_0^\prime\\t_0^\prime \end{array} \right) =U_u^{wb}\left(\begin{array}{c} 
u_0 \\c_0 \\t_0 \end{array} \right),
\hspace{.5cm}
\left(\begin{array}{c} d_0^\prime \\s_0^\prime \\b_0^\prime \end{array} \right) =U_d^{bf}\left(\begin{array}{c} 
d_0 \\s_0 \\b_0 \end{array} \right),
\end{equation}
where $U_u^{wb}$ and $U_d^{bf}$ are complex diagonal matrices as presented in appendix (\ref{bB}).

So, in the primed basis, the algebra reduces to diagonalize the two real symmetric mass matrices (\ref{wbr}) for $q=u$ 

and (\ref{bfr}) for $q=d$, which are:

\[M_u^{I} =\left(\begin{array}{ccc} 

0 & |a_u| & 0 \\ 

|a_u| & c_u & |b_u|  \\

0 & |b_u|  & d_u

\end{array}\right),\hspace{0.2cm}
M_d^{I} =\left(\begin{array}{ccc} 

0 & |a_d| & 0 \\ 

|a_d| & c_d & 0  \\

0 & 0  & d_d

\end{array}\right),\]

diagonalization carried through in Appendix (\ref{bB}).

From the former analysis we can evaluate the mixing matrix $V_{CKM}^{I}=O_u^{wb}U_u^{wb}U_d^{bf\dagger} [O_d^{bf}]^T$, where $U_u^{wb}$ and $U_d^{bf}$ are as defined in (\ref{upph}) and calculated in Appendix (\ref{bB}) and $O^{wb}_u$ and $O^{bf}_d$ are given by (\ref{rotwb}) and (\ref{obf}) respectively.

The elements of the CKM mixing matrix can be expressed in the context of this pattern as

\begin{eqnarray}\nonumber
(V_{CKM}^{I})_{lm}&=&(O_u^{wb})_{l1}(O_d^{bf})_{m1}+e^{i\phi_1}(O_u^{wb})_{l2}(O_d^{bf})_{m2}\\ \label{vckI}
&&+e^{i\phi_2}(O_u^{wb})_{l3}(O_d^{bf})_{m3},
\end{eqnarray}

where $\phi_1=(\alpha_u^w-\alpha_d^b)$ and $\phi_2=(\alpha_u^w+\beta_u^w)$.

The use of the matrices in appendix (\ref{bB}) allow us to write the following analytic mixing matrix:

{\begin{widetext}

\[V_{CKM}^I=\left(
\begin{array}{ccc}
\frac{\sqrt{d_u-m_u} \left(\sqrt{m_c m_s m_t}+e^{i\phi_1} \sqrt{d_u m_d m_u}\right)}
{\sqrt{d_u(m_d+m_s)(m_t-m_u) (m_c+m_u)}} 
& \frac{\sqrt{d_u-m_u} \left(-\sqrt{m_c m_d m_t}+e^{i\phi_1} \sqrt{d_um_sm_u}\right)}
{\sqrt{d_u(m_d+m_s)(m_t-m_u)(m_c+m_u),}}
& -e^{i\phi_2}\frac{\sqrt{(d_u + m_c) (m_t-d_u) m_u}}{\sqrt{d_u (m_t-m_u) (m_c+m_u)}} \\
\frac{-e^{i\phi_1} \sqrt{d_u m_c m_d (d_u+m_c)}+\sqrt{(d_u+m_c) m_s m_t m_u}}
{\sqrt{d_u(m_d+m_s) (m_c+m_t) (m_c+m_u)}} 
& -\frac{e^{i\phi_1} \sqrt{d_u m_c m_s}+\sqrt{m_d m_t m_u}}{\sqrt{\frac{d_u (m_d+ m_s) (m_c+m_t) (m_c+m_u)}{d_u+m_c}}} 
& e^{i\phi_2}\frac{\sqrt{m_c(m_t-d_u)(d_u-m_u)}}{\sqrt{d_u(m_c+m_t)(m_c+m_u)}}  \\
\frac{\sqrt{-d_u+ m_t} \left(e^{i\phi_1} \sqrt{d_u m_d m_t}+\sqrt{m_c m_s m_u}\right)}
{\sqrt{d_u(m_d+m_s) (m_c+m_t) (m_t-m_u)}} 
& \frac{\sqrt{-d_u+m_t} \left(e^{i\phi_1} 
\sqrt{d_um_sm_t}-\sqrt{m_cm_dm_u}\right)}{\sqrt{d_u (m_d+m_s) (m_c+m_t) (m_t-m_u)}}
& e^{i\phi_2}\frac{\sqrt{(d_u+m_c)m_t (d_u-m_u)}}{\sqrt{d_u (m_c+m_t) (m_t-m_u)}}
\end{array}
\right),\]\\

\end{widetext}}

\noindent

where as anticipated, the last column is a function of the parameters in the up quark mass matrix only, and a common phase

$e^{i\phi_2}$ in that column can be removed by a redefinition of the $b$ quark field.

Our approach now is the following: first we use the free parameter $d_u$ in order to fit the experimental value $V_{cb}$ 

and then use the phase $\phi_1$ to fit $|V_{us}|$; when the central quark mass values are used we obtain for this pattern 

the result $d_u\approx m_t$ and $\phi_1\approx 1.6$. The next step is to fine tune both $d_u$ and $\phi_1$ using a random 

numerical analysis, using quark masses $m_u$ and $m_d$ at $2\sigma$ values with the other four quark masses at 

$1\sigma$~\cite{Qexp}), aiming to get results as close as possible to the ones presented in $V^{(exp)}$ in (\ref{maexp}). 

Finally, for the best values obtained we calculate the CP asymmetries $\alpha,\;\beta$ and $\gamma$ as in~(\ref{cpangl}).

Notice that a crude approximation for which $d_u\equiv m_t$, using further $m_t>>m_c>>m_u$, implies from the mixing matrix 

$V_{CKM}^I$ that

\begin{equation}\label{vucapx}
V_{us}\approx \sqrt{\frac{m_s}{m_s+m_d}}\left|\sqrt{\frac{m_d}{m_s}}-e^{i\phi}\sqrt{\frac{m_u}{m_c}}\right|,
\end{equation}

a form advocated in several references~\cite{kim}. But to set $d_u=m_t$ implies $b_u=0$ according to Eq. (\ref{cq4}), and 

not only we get back to a parallel six texture zero pattern with $\theta_{13}=\theta_{23}=0$ (something that can be seen also 

from $V_{CKM}^I$, the $3\times 3$ mixing matrix above). So, our fine tunning approach is mandatory, $d_u$ must be different 

from $m_t$ and any analysis done for which expression (\ref{vucapx}) plays a central role, must produce dubious predictions. 

In this regard, the conclusions in Ref.~\cite{kim}, which are contrary to ours, may be wrong.

The numerical subroutines used throws for pattern I the following numbers (mass parameters are in GeV's and angles in radians):

\[m_u=0.0023, \;\; m_c=0.560, \;\; m_t=172, \]

\[m_d=0.0029, \;\; m_s=0.06 , \;\; m_b=2.89, \]

\[d_u=171.721<m_t,\; \;\; \phi_1=1.61;\]

\noindent

which imply the following $3\times 3$ mixing matrix:

\[\left(
\begin{array}{ccc}

 0.97428 & 0.22532  & 0.00264 \\

 0.22517 & 0.97349  & 0.04020  \\

 0.00865 & 0.03934  & 0.99919

\end{array}
\right),\]

numbers in quite good agreement with the experimental measured values. Finally, the three angles of the unitary triangle 

of the B decays CP asymmetries calculated according to (\ref{cpangl}) for this pattern are:

\[ (\alpha ,\; \beta ,\; \gamma)_{th}^{I}= (90.79,\; 16.51, \; 72.68), \]

\noindent

which not only close the triangle, but are such that $\alpha$ and $\gamma$ agree with the measured value at $1\sigma$ and 

$\beta$ at $2\sigma$.

\subsubsection{Pattern II}

Notice for this pattern that the mixing angles $\theta_{13}$ and $\theta_{23}$ between the third family and the first two 

ones come only from the down quark sector. Also, this pattern is obtained from the previous one by the replacements: 

$M_d^{I}\longrightarrow M_u^{II}$ and $M_u^{I}\longrightarrow M_d^{II}$ and $d_u\rightarrow d_d$, where $d_d$ is the free 

parameter of this pattern.

As in the previous pattern, we start by removing the complex phases from $M_u^{II}$ and $M_d^{II}$ and then use the results 

in appendix B in order to diagonalize the orthogonal mass matrices. When done, the CKM mixing matrix for this pattern reads: 

$V_{CKM}^{II}=O_u^{bf}U_u^{bf}U_d^{wb\dagger} [O_d^{wb}]^T$, where the orthogonal and unitary matrices are presented in appendix 

(B). The elements of the CKM mixing matrix can be expressed now as

\begin{eqnarray}\nonumber
(V_{CKM}^{II})_{lm}&=&(O_u^{bf})_{l1}(O_d^{wb})_{m1}+e^{i\phi_1}(O_u^{bf})_{l2}(O_d^{wb})_{m2}\\ \label{vckII}
&&+e^{i\phi_2}(O_u^{bf})_{l3}(O_d^{wb})_{m3},
\end{eqnarray}

where $\phi_1=(\alpha_u^b-\alpha_d^w)$ and $\phi_2=-(\alpha_d^w+\beta_d^w)$, which allows us to write the following analytic 

forms:

\begin{widetext}

\begin{equation}
\begin{split}
V_{ud}^{II}&=\frac{ \sqrt{m_cm_b \left(d_d-m_d\right) m_s}}{\sqrt{\left(m_c+m_u\right)d_d \left(m_b-m_d\right) \left(m_d+m_s\right)}}+
\frac{e^{i\phi_1} \sqrt{m_u\left(d_d-m_d\right) m_d} }{\sqrt{\left(m_c+m_u\right)\left(m_b-m_d\right) \left(m_d+m_s\right)}},\\
V_{us}^{II}&=\frac{\sqrt{m_cm_b m_d \left(d_d+m_s\right)}}{\sqrt{\left(m_c+m_u\right)d_d\left(m_b+m_s\right) \left(m_d+m_s\right)}}-
\frac{e^{i\phi_1} \sqrt{m_s m_u\left(d_d+m_s\right)} }{\sqrt{\left(m_c+m_u\right)\left(m_b+m_s\right) \left(m_d+m_s\right)}},\\
V_{ub}^{II}&=\frac{ \sqrt{\left(-d_d+m_b\right) m_d m_sm_c}}{\sqrt{\left(m_c+m_u\right)d_d\left(m_b-m_d\right) \left(m_b+m_s\right)}}+
\frac{e^{i\phi_1} \sqrt{m_b \left(-d_d+m_b\right)m_u} }{\sqrt{\left(m_c+m_u\right)\left(m_b-m_d\right) \left(m_b+m_s\right)}},\\
V_{cd}^{II}&=\frac{e^{i\phi_1}\sqrt{\left(d_d-m_d\right) m_dm_c}}{\sqrt{\left(m_c+m_u\right)\left(m_b-m_d\right) \left(m_d+m_s\right)}}-\frac{\sqrt{m_um_b
\left(d_d-m_d\right) m_s}}{\sqrt{\left(m_c+m_u\right)d_d \left(m_b-m_d\right) \left(m_d+m_s\right)}},\\
V_{cs}^{II}&=-\frac{e^{i\phi_1} \sqrt{m_sm_c \left(d_d+m_s\right)}}{\sqrt{\left(m_c+m_u\right)\left(m_b+m_s\right) 
\left(m_d+m_s\right)}}-\frac{\sqrt{m_bm_d \left(d_d+m_s\right)m_u} }{\sqrt{\left(m_c+m_u\right)d_d \left(m_b+m_s\right) 
\left(m_d+m_s\right)}},\\
V_{cb}^{II}&=\frac{e^{i \phi_1}\sqrt{m_cm_b \left(-d_d+m_b\right)}}{\sqrt{\left(m_c+m_u\right)\left(m_b-m_d\right) 
\left(m_b+m_s\right)}}-\frac{\sqrt{\left(-d_d+m_b\right)
m_um_d m_s} }{\sqrt{\left(m_c+m_u\right)d_d\left(m_b-m_d\right) \left(m_b+m_s\right)}},\\
V_{td}^{II}&=-e^{i\phi_2} \sqrt{\frac{\left(m_b-d_d\right) m_d \left(d_d+m_s\right)}{d_d \left(m_b-m_d\right) \left(m_d+m_s\right)}},\\
V_{ts}^{II}&=e^{i\phi_2} \sqrt{\frac{\left(m_b-d_d\right) \left(d_d-m_d\right) m_s}{d_d \left(m_b+m_s\right) \left(m_d+m_s\right)}},\\ 
V_{tb}^{II}&=e^{i\phi_2} \sqrt{\frac{m_b \left(d_d-m_d\right) \left(d_d+m_s\right)}{d_d \left(m_b-m_d\right) \left(m_b+m_s\right)}},\\ 
\end{split}
\end{equation}

\end{widetext}

where again the last row is a function of the parameters in the down quark mass matrix only, and the phase 

$e^{i\phi_2}$ can be removed by a redefinition of the field for the quark $t$. The form of the former mixing entries 

recommends to use $d_d$ to fit the experimental measured value $V_{ts}$ and then $\phi_1$ to fit $V_{us}$, which 

produces $d_d\approx m_b$ and $\phi_1\approx 1.45$.

Notice that for $d_d=m_d$ and $m_c\gg m_u$, $V_{us}^{II}$ acquires the form presented in Eq.~({\ref{vucapx}). 

But again, $d_d=m_b$ drives  pattern II to a six texture zero ``ansatz'' with $\theta_{13}=\theta_{23}=0$, implying 

once more a mandatory fine tunning for the free parameter $d_d$.

The random numerical analysis for pattern II throws the following numbers (mass parameters in GeV's and angles in radians 

as before):

\[m_u=0.0023, \;\; m_c=0.560, \;\; m_t=172,\]

\[m_d=0.0029, \;\; m_s=0.060,  \;\; m_b=2.89,\]

\[d_d=2.8855<m_b,\;\;\;\; \phi_1=1.6,\]

which imply the following $3\times 3$ mixing matrix:

\[\left(
\begin{array}{ccc}

 0.97437 & 0.22493  & 0.00250 \\

 0.22478 & 0.97363  & 0.03900  \\

 0.00856 & 0.03813  & 0.99924

\end{array}
\right),\]

numbers again in good agreement with the experimental measured values. Finally, the three angles of the unitary triangle 

of the B decays CP asymmetries are calculated for this pattern to be:

\[ (\alpha ,\; \beta ,\; \gamma)_{th}^{(II)}= (86.78,\; 16.11, \; 77.11), \]

which again close the triangle and are such that $\alpha$ and $\gamma$ agree with the measured value at $1\sigma$ with a 

value for $\beta$ at  $2\sigma$.

\subsubsection{Pattern III}

Proceeding as in the two previous patterns and using the notation introduced in appendix (B) we have now  $V_{CKM}^{III}=O_u^{nn}U_u^{nn}U_d^{wb\dagger} [O_d^{wb}]^T$, where the elements of the CKM mixing matrix can be expressed as

\begin{eqnarray}\nonumber
(V_{CKM}^{III})_{lm}&=&(O_u^{nn})_{l1}(O_d^{wb})_{m1}+e^{i\phi_1}(O_u^{nn})_{l2}(O_d^{wb})_{m2}\\ \label{vckIII}
&&+e^{i\phi_2}(O_u^{nn})_{l3}(O_d^{wb})_{m3},
\end{eqnarray}

where $\phi_1=(\alpha_u^n-\alpha_d^w)$ and $\phi_2=(\alpha_u^n-\alpha_d^w+\beta_u^n-\beta_d^w)$, which allows us to write the following analytic forms:

{\small \begin{widetext}

\begin{equation}
\begin{split}
V_{ud}^{III}&=\sqrt{ \frac{(m_t-m_c)}{(m_t-m_u)(m_c+m_u)(m_b-m_d)(m_d-m_s)}}
\left[\sqrt{\frac{m_tm_cm_bm_s(d_d-m_d)}{d_d(m_t-m_c+m_u)}}\right.\\
&\left.+e^{i\phi_1}\sqrt{m_um_d(d_d-m_d)} + e^{i\phi_2}\sqrt{\frac{m_um_d(m_t+m_u)(m_c-m_u)(m_b-d_d)(d_d+m_s)}{d_d(m_t-m_c+m_u)(m_t-m_c)}}\right],\\
V_{us}^{III}&=\sqrt{ \frac{(m_t-m_c)}{(m_t-m_u)(m_c+m_u)(m_b+m_s)(m_d+m_s)}}
\left[\sqrt{\frac{m_tm_cm_bm_d(d_d+m_s)}{d_d(m_t-m_c+m_u)}}\right.\\
&-\left.e^{i\phi_1}\sqrt{m_um_s(d_d+m_s)} - e^{i\phi_2}\sqrt{\frac{m_um_s(m_t+m_u)(m_c-m_u)(m_b-d_d)(d_d+m_d)}{d_d(m_t-m_c+m_u)(m_t-m_c)}}\right],\\
V_{ub}^{III}&=\sqrt{ \frac{(m_t-m_c)}{(m_t-m_u)(m_c+m_u)(m_b+m_s)(m_b+m_d)}}
\left[\sqrt{\frac{m_tm_cm_sm_d(m_b-d_d)}{d_d(m_t-m_c+m_u)}}\right.\\
&+\left.e^{i\phi_1}\sqrt{m_um_b(m_b-d_d)} - e^{i\phi_2}\sqrt{\frac{m_um_b(m_t+m_u)(m_c-m_u)(d_d+m_s)(d_d-m_d)}{d_d(m_t-m_c+m_u)(m_t-m_c)}}\right],\\
V_{cd}^{III}&=\sqrt{ \frac{(m_t+m_u)}{(m_t+m_c)(m_c+m_u)(m_b-m_d)(m_d+m_s)}}
\left[\sqrt{\frac{m_tm_um_bm_s(d_d-m_d)}{d_d(m_t-m_c+m_u)}}\right.\\
&-\left.e^{i\phi_1}\sqrt{m_cm_d(d_d-m_d)} - e^{i\phi_2}\sqrt{\frac{m_cm_d(m_t-m_c)(m_c-m_u)(m_b-d_d)(d_d+m_s)}{d_d(m_t-m_c+m_u)(m_t+m_u)}}\right],\\
V_{cs}^{III}&=\sqrt{ \frac{(m_t+m_u)}{(m_t+m_c)(m_c+m_u)(m_b+m_s)(m_d+m_s)}}
\left[\sqrt{\frac{m_tm_um_bm_d(d_d+m_s)}{d_d(m_t-m_c+m_u)}}\right.\\
&+\left.e^{i\phi_1}\sqrt{m_cm_s(d_d+m_s)} + e^{i\phi_2}\sqrt{\frac{m_cm_s(m_t-m_c)(m_c-m_u)(m_b-d_d)(d_d-m_d)}{d_d(m_t-m_c+m_u)(m_t+m_u)}}\right],\\
V_{cb}^{III}&=\sqrt{ \frac{(m_t+m_u)}{(m_t+m_c)(m_c+m_u)(m_b+m_s)(m_b-m_d)}}
\left[\sqrt{\frac{m_tm_um_sm_d(m_b-d_d)}{d_d(m_t-m_c+m_u)}}\right.\\
&-\left.e^{i\phi_1}\sqrt{m_cm_b(m_b-d_d)} + e^{i\phi_2}\sqrt{\frac{m_cm_b(m_t-m_c)(m_c-m_u)(d_d-m_d)(d_d+m_s)}{d_d(m_t-m_c+m_u)(m_t+m_u)}}\right],\\
V_{td}^{III}&=\sqrt{ \frac{(m_c-m_u)}{(m_t+m_c)(m_t-m_u)(m_s+m_d)(m_b-m_d)}}
\left[\sqrt{\frac{m_cm_um_bm_s(d_d-m_d)}{d_d(m_t-m_c+m_u)}}\right.\\
&+\left.e^{i\phi_1}\sqrt{m_tm_d(d_d-m_d)} - e^{i\phi_2}\sqrt{\frac{m_tm_d(m_t-m_c)(m_t+m_u)(m_b-d_d)(d_d+m_s)}{d_d(m_t-m_c+m_u)(m_c-m_u)}}\right],\\
V_{ts}^{III}&=\sqrt{ \frac{(m_c-m_u)}{(m_t+m_c)(m_t-m_u)(m_b+m_s)(m_s+m_d)}}
\left[\sqrt{\frac{m_cm_um_bm_d(d_d+m_s)}{d_d(m_t-m_c+m_u)}}\right.\\
&-\left.e^{i\phi_1}\sqrt{m_tm_s(d_d+m_s)} + e^{i\phi_2}\sqrt{\frac{m_tm_s(m_t-m_c)(m_t+m_u)(m_b-d_d)(d_d-m_d)}{d_d(m_t-m_c+m_u)(m_c-m_u)}}\right],\\
V_{tb}^{III}&=\sqrt{ \frac{(m_c-m_u)}{(m_t+m_c)(m_t-m_u)(m_b+m_s)(m_b-m_d)}}
\left[\sqrt{\frac{m_cm_um_dm_s(m_b-d_d)}{d_d(m_t-m_c+m_u)}}\right.\\
&+\left.e^{i\phi_1}\sqrt{m_tm_b(m_b-d_d)} + e^{i\phi_2}\sqrt{\frac{m_tm_b(m_t-m_c)(m_t+m_u)(d_d+m_s)(d_d-m_d)}{d_d(m_t-m_c+m_u)(m_c-m_u)}}\right].\\
\end{split}
\end{equation}

\end{widetext}}

Notice that for $d_d=m_d$ and $m_t\gg m_c\gg m_u$, $V_{us}^{III}$ acquires the form presented in Eq.~({\ref{vucapx}). 

But again, $d_d=m_b$ drives  pattern III to a six texture zero ``ansatz''.

Written in the previous form, all the entries for $V_{CKM}^{III}$ includes two different phases, $\phi_1$ and $\phi_2$, 

none of them absorbed immediately by a redefinition of a single quark field; but since it is a well known fact that the 

SM mixing matrix can be parametrized with only one single phase, our analysis makes sense only for the following three 

different cases, which must be studied separately:

\begin{itemize}

 \item Case 1: $\phi_1\neq 0,\;\;\phi_2=0$.

 \item Case 2: $\phi_1 = 0,\;\;\phi_2\neq 0$.

 \item Case 3: $\phi_1=\phi_2\neq 0$.

\end{itemize}

The numerical analysis for pattern III and for $\phi_1=\phi_2$ throws the following numbers (mass parameters in GeV's and 

angles in radians as before):

\[m_u=0.0023, \;\; m_c=0.560, \;\; m_t=172,\]

\[m_d=0.0029, \;\; m_s=0.06,  \;\; m_b=2.89,\]

\[d_d=2.889,\;\;\;\; \phi_1=\phi_2=1.6,\]

which imply the following $3\times 3$ mixing matrix:

\[\left(
\begin{array}{ccc}

 0.97437 & 0.22493  & 0.00247 \\

 0.22408 & 0.97365  & 0.03837  \\

 0.00817 & 0.03757  & 0.99926

\end{array}
\right),\]

numbers again in good agreement with the experimental measured values. Finally, the three angles of the unitary triangle of the B decays CP asymmetries calculated for this Pattern are:

\[ (\alpha ,\; \beta ,\; \gamma)_{th}^{(III)}= (92.73,\; 16.22, \; 71.06), \]

which again close the triangle and are such that $\alpha$ and $\gamma$ agree with the measured value at $1\sigma$ and $\beta$ at $2\sigma$.

\subsubsection{Pattern IV}

Proceeding in a similar way we have for this pattern that the CKM mixing matrix reads now $V_{CKM}^{IV}=O_u^{ss}U_u^{ss}U_d^{bf\dagger} [O_d^{bf}]^T$, and the elements of the CKM mixing matrix can be expressed as

\begin{eqnarray}\nonumber
(V_{CKM}^{IV})_{lm}&=&(O_u^{ss})_{l1}(O_d^{bf})_{m1}+e^{i\phi_1}(O_u^{ss})_{l2}(O_d^{bf})_{m2}\\ \label{vckIV}
&&+e^{i\phi_2}(O_u^{ss})_{l3}(O_d^{bf})_{m3},
\end{eqnarray}

where $\phi_1=(\alpha_u^d-\alpha_d^w)$ and $\phi_2=(\alpha_u^d+\beta_u^w-\alpha_d^w)$, which allows us to write the following 

analytic forms:

\begin{widetext}

{\small\begin{equation}\label{pattern4}
\begin{split}
V_{ud}^{IV}&=\frac{\sqrt{m_tm_c m_s} \left(m_t-d_u+m_c\right)+e^{i\phi_1}\sqrt{m_d m_u (d_u-m_c+m_u)(m_t+m_c-d_u)(m_t-m_u-d_u)}}{\sqrt{(m_d+m_s)(-d_u+m_c+m_t)(m_c+m_u)(-d_u+m_c+m_t-m_u)(m_t+m_u)}},\\
V_{us}^{IV}&=\frac{-\sqrt{m_tm_c m_d} \left(m_t-d_u+m_c\right)+e^{i\phi_1}\sqrt{m_s m_u (d_u-m_c+m_u)(m_t+m_c-d_u)(m_t-m_u-d_u)}}{\sqrt{(m_d+m_s)(-d_u+m_c+m_t)(m_c+m_u)(-d_u+m_c+m_t-m_u)(m_t+m_u)}},\\
V_{ub}^{IV}&=e^{i\phi_2}\frac{\sqrt{m_u\left(m_t-d_u+m_c\right)}}{\sqrt{\left(m_c+m_u\right) \left(m_t+m_u\right)}},\\
V_{cd}^{IV}&=\frac{-e^{i\phi_1} m_c \sqrt{m_d (m_t-d_u+m_c)(d_u-m_c+m_u)}+\sqrt{m_sm_cm_um_t(m_t-d_u-m_u)}}{\sqrt{m_c(m_d+m_s)(m_c+m_u)(m_t-m_c)(m_t+m_c-m_u-d_u)}},\\
V_{cs}^{IV}&=-\frac{e^{i\phi_1} m_c \sqrt{m_s (m_t-d_u+m_c)(d_u-m_c+m_u)}+\sqrt{m_dm_cm_um_t(m_t-d_u-m_u)}}{\sqrt{m_c(m_d+m_s)(m_c+m_u)(m_t-m_c)(m_t+m_c-m_u-d_u)}},\\
V_{cb}^{IV}&=e^{i\phi_2}\frac{\sqrt{m_c\left(m_t-d_u+m_u\right)}}{\sqrt{\left(m_t-m_c\right) \left(m_c+m_u\right)}},\\
V_{td}^{IV}&=\frac{e^{i\phi_1} \sqrt{m_d m_t (m_t-d_u+m_c)(d_u-m_c+m_u)(m_t-d_u-m_u)}+\sqrt{m_sm_cm_u}(d_u-m_c+m_u)}{\sqrt{(m_d+m_s)(d_u-m_c+m_u)(m_t+m_c-m_u-d_u)(m_t-m_c)(m_t+m_u)}},\\
V_{ts}^{IV}&=\frac{e^{i\phi_1} \sqrt{m_s m_t (m_t-d_u+m_c)(d_u-m_c+m_u)(m_t-d_u-m_u)}+\sqrt{m_dm_cm_u}(-d_u+m_c-m_u)}{\sqrt{(m_d+m_s)(d_u-m_c+m_u)(m_t+m_c-m_u-d_u)(m_t-m_c)(m_t+m_u)}},\\
V_{tb}^{IV}&=e^{i\phi_2}\frac{\sqrt{m_t\left(d_u-m_c+m_u\right)}}{\sqrt{\left(m_t-m_c\right) \left(m_t+m_u\right)}},\\
\end{split}
\end{equation}}

\end{widetext}

where again the common phase $e^{i\phi_2}$ in the last column can be removed by a redefinition of the $b$ quark field.

The numerical analysis for pattern IV are: (mass parameters in GeV's and angles in radians as before):

\[m_u=0.00127, \;\; m_c=0.619, \;\; m_t=171.7,\]

\[m_d=0.0028, \;\; m_s=0.055,  \;\; m_b=2.89,\]

\[d_u=171.39,\;\;\;\; \phi_1=1.9,\]

which imply the following $3\times 3$ mixing matrix:

\[\left(
\begin{array}{ccc}

 0.97331 & 0.22549  & 0.00333 \\

 0.22934 & 0.97242  & 0.04144  \\

 0.00889 & 0.04163  & 0.99909

\end{array}
\right),\]

numbers again in good agreement with the experimental measured values.

Finally, the three angles of the unitary triangle of the B decays CP asymmetries calculated for this Pattern are:

\[ (\alpha ,\; \beta ,\; \gamma)_{th}^{(IV)}= (95.23,\; 19.36, \; 65.42), \]

which not only close the triangle, but the three agree with the measured value at $1\sigma$.

\subsubsection{Pattern V}

Proceeding in a similar way we have for this pattern that the CKM mixing matrix reads now 

$V_{CKM}^{V}=O_u^{dd}U_u^{dd}U_d^{wb\dagger} [O_d^{wb}]^T$, and the elements of the CKM mixing matrix can be expressed as

\begin{eqnarray}\nonumber
(V_{CKM}^{V})_{lm}&=&(O_u^{dd})_{l1}(O_d^{wb})_{m1}+e^{i\phi_1}(O_u^{dd})_{l2}(O_d^{wb})_{m2}\\ \label{vckV}
&&+e^{i\phi_2}(O_u^{dd})_{l3}(O_d^{wb})_{m3},
\end{eqnarray}

where $\phi_1=-\alpha_d^w$ and $\phi_2=(\alpha_u^d-\alpha_d^w-\beta_d^w)$, which allows us to write the following analytic forms:

\begin{widetext}

\begin{equation}
\begin{split}
V_{ud}^{V}&=\frac{\sqrt{m_b m_s m_t \left(d_d-m_d\right)}- e^{i\phi_2} \sqrt{(m_b-d_d)\left(d_d+m_s\right)m_d m_u}}
{\sqrt{d_d\left(m_b-m_d\right) \left(m_d+m_s\right) \left(m_t+m_u\right)}},\\
V_{us}^{V}&=\frac{\sqrt{m_b m_d m_t\left(d_d+m_s\right)}+ e^{i\phi_2} \sqrt{(m_b-d_d)\left(d_d-m_d\right)m_s m_u}}
{\sqrt{\left(m_t+m_u\right)d_d \left(m_b+m_s\right) \left(m_s+m_d\right)}},\\
V_{ub}^{V}&=\frac{\sqrt{m_dm_sm_t \left(m_b-d_d\right)}+ e^{i\phi_2} \sqrt{m_bm_u(m_s+d_d)\left(d_d-m_d\right)}}
{\sqrt{\left(m_t+m_u\right)d_d \left(m_b-m_d\right) \left(m_b+m_s\right)}},\\
V_{cd}^{V}&=e^{i\phi_1}\frac{\sqrt{m_d(d_d-m_d})}{\sqrt{(m_b-m_d)(m_d+m_s})},\\
V_{cs}^{V}&=-e^{i\phi_1}\frac{\sqrt{m_s(d_d+m_s})}{\sqrt{(m_b+m_s)(m_d+m_s})},\\
V_{cb}^{V}&=e^{i\phi_1}\frac{\sqrt{m_b(m_b-d_d})}{\sqrt{(m_b-m_d)(m_b+m_s})},\\
V_{td}^{V}&=\frac{-\sqrt{m_um_bm_s \left(d_d-m_d\right)}- e^{i\phi_2} \sqrt{m_tm_d(m_b-d_d)\left(d_d+m_s\right)}}
{\sqrt{\left(m_t+m_u\right)d_d \left(m_b-m_d\right) \left(m_s+m_d\right)}},\\
V_{ts}^{V}&=\frac{-\sqrt{m_um_bm_d \left(d_d+m_s\right)}+ e^{i\phi_2} \sqrt{m_tm_s(m_b-d_d)\left(d_d-m_d\right)}}
{\sqrt{\left(m_t+m_u\right)d_d \left(m_b+m_s\right) \left(m_s+m_d\right)}},\\
V_{tb}^{V}&=\frac{-\sqrt{m_um_dm_s \left(m_b-d_d\right)}+ e^{i\phi_2} \sqrt{m_tm_b(d_d+m_s)\left(d_d-m_d\right)}}
{\sqrt{\left(m_t+m_u\right)d_d \left(m_b+m_s\right) \left(m_b-m_d\right)}},\\
\end{split}
\end{equation}

\end{widetext}

where the phase $\phi_1$  can be removed by a redefinition of the charm quark field.

The numerical analysis throws now the following numbers (mass parameters in GeV's and angles in radians as before):

\[m_u=0.00127, \;\; m_c=0.619, \;\; m_t=171.7,\]

\[m_d=0.0029, \;\; m_s=0.055,  \;\; m_b=2.89,\]

\[d_d=2.885,\;\;\;\; \phi_2=1.9,\]

which imply the following $3\times 3$ mixing matrix:

\[\left(
\begin{array}{ccc}

 0.97463 & 0.22377  & 0.00266 \\

 0.22361 & 0.97381  & 0.04122  \\

 0.00891 & 0.04034  & 0.99915

\end{array}
\right),\]

numbers again in quite good agreement with the experimental measured values. Finally, the three angles of the unitary 

triangle of the B decays CP asymmetries calculated for this pattern are:

\[ (\alpha ,\; \beta ,\; \gamma)_{th}^{(V)}= (88.83,\; 16.36, \; 74.81), \]

which close the triangle and are such that $\alpha$ and $\gamma$ agree with the measured value at $1\sigma$, with a value 

for $\beta$ at $2\sigma$.

\subsubsection{Related patterns}

The analysis shows that a weak basis transformation using the unitary matrix

\[U=\left(\begin{array}{ccc}

           1 & 0 & 0 \\

           0 & 0 & 1\\

           0 & 1 & 0

          \end{array}\right) = U^\dagger,\]

shows that $UM^{ss}U=M^{wb}$ and also that $UM^{bf}U=M^{dd}$. So, a pattern like

\[M_u^{VI}=\left(\begin{array}{ccc}

0     &  0   & a_u     \\

0     & b_u  & 0       \\

a_u^* & 0    & c_u

\end{array}\right);\;\;\;\;\;
M_d^{VI}=\left(\begin{array}{ccc}

0     &  0   & a_d     \\

0     & d_d  & b_d  \\

a_d^* & b_d^* & c_d
\end{array}\right)\]

should be physically equivalent to Pattern II.

A pattern like

\[M_u^{VII}=\left(\begin{array}{ccc}

0     &  a_u   & 0     \\

a_u^* & c_u    & b_u   \\

0     & b_u^*  & d_u

\end{array}\right);\;\;\;\;\;
M_d^{VII}=\left(\begin{array}{ccc}

0     &  0   & a_d     \\

0     & c_d  & 0       \\

a_d^* & 0    & d_d

\end{array}\right)\]

should be physically equivalent to Pattern IV.

And a pattern like

\[M_u^{VIII}=\left(\begin{array}{ccc}

0     &  a_u & 0    \\

a_u^* & b_u  & 0    \\

0     & 0    & c_u

\end{array}\right);\;\;\;\;\;
M_d^{VIII}=\left(\begin{array}{ccc}

0     &  0   & a_d     \\

0     & c_d  & b_d  \\

a_d^* & b_d^* & d_d

\end{array}\right)\]

should be physically equivalent to Pattern V.

\section{\label{sec:sec5}Discussion of the results}

All the five texture zeros patterns studied here, produce numerical results for the CKM mixing matrix and CP violating parameters $(\alpha,\;\beta,\;\gamma)$ in agreement with the measured values up to $2\sigma$. Pattern IV reproduces them up to $1\sigma$. Since the results are presented for the best fit values, all the patterns can be taken predictive, as far as the up to date measured numbers are concerned.

But: how sensitive is our numerical analysis to the variations of the quark mass values? Since the answer to this question is pattern dependent, let us do it for pattern IV, the most predictive one.

Since $m_t\gg m_c\gg m_u$ and the experimental uncertainty in the top quark mass $\Delta m_t$ is such that $\Delta m_t>m_c>m_u$; then for the up quark sector it is enough to consider variations to the top quark mass in the interval $168.7\leq m_t{\mbox (GeV)}\leq 174.7$.

Our results show that the numerical predicted values remain stable for the entire interval, as far as the difference $(m_t-d_u)$ stays constant, in a range from 0.28 to 0.33 [notice that all the CKM entries for pattern IV but $V_{tb}^{IV}\sim 1$ in Eq.(\ref{pattern4}), depend upon the difference $(m_t-d_u)$].

Now, for the quark masses in the down sector, the CKM mixing matrix for pattern IV does not depend upon the $b$ quark mass (due to the block form of the down quark mass matrix in this particular pattern). For the other down quark masses we have that the numerical results are stable as far as the strange quark mass is in the range $0.051\leq m_s\mbox{(GeV)}\le0.060$ and for the down quark mass $0.025\leq m_d\mbox{(GeV)}\le0.030$.

Can we relate our different parametrization of the CKM mixing matrix with other different parameterization? In special to the advocated standard parametrization~\cite{smpar} in matrix (\ref{vdspar})?, the answer to this  question is an academic exercise without much physical sense, because the adoption of a parametrization of the flavor mixing is an arbitrary fact and not a physical issue. In particular, our phase $\phi_1$ in pattern IV and the phase $\delta$ of the standard parametrization, are not physical quantities. The real physical parameters are in this case the CP violating mixing angles $(\alpha,\;\beta,\;\gamma)$.

Also, the rotation parameters $\theta_{12}$, $\theta_{23}$ and $\theta_{13}$ of the standard parameterization, are not physical entities either. 

For example, in the standard parameterization, $\theta_{12}=\tan^{-1}(V_{us}/V_{ud})$ which is a real number. In any of our parametrization, $(V_{us}/V_{ud})$ is a complex number, without direct relation to $\theta_{12}$.

Since in general grounds one expects that the flavor mixing parameters do depend on the elements of the quark mass matrices, we feel that our CKM mixing entries are more realistic that any ad-hoc parametrization.

Can we do the mathematical exercise to relate the phase $\phi_1$ in pattern IV to the phase $\delta$ of the standard parametrization? (or to any other phase of an arbitrary parametrization?). The answer is yes and the simplest way to do it is by the use of the Jarlskog~\cite{jarlk} invariant, which by the sake of convenience we use as:

\begin{equation}\label{jarl}
 J=\pm \mbox{Im.} (V_{us}V_{cb}V^*_{ub}V^*_{cs}),
\end{equation}

which for the standard parametrization takes the form: $J=c_{12}c_{13}^2c_{23}s_{12}s_{13}s_{23}\sin\delta$, with a numerical evaluation producing~\cite{pdg}

$(2.80\leq J\leq 3.16)\times 10^{-5}$.

For pattern IV, an analytic expression for $J$ can be obtained, but it is prohibitive long to quote it here. A numerical evaluation is more viable, and for the best fit values for the quark masses we get $J=2.87\times 10^{-5}$ which falls in the allowed range of the standard parametrization. Of course, this value is sensitive to the variation of the quark mass values. In particular we see that an strange quark mass (keeping the other masses in the best fit values) in the interval $0.048\leq m_s(\mbox{GeV})\leq 0.056$ is compatible with the range of the Jarlskog invariant of the standard parametrization.

\section{\label{sec:sec6}Conclusions}

In this paper we have presented our exhaustive study of all the possible five texture zeros quark mass matrices and their 

connection with the CP violation phenomena for three SM families. Analytic and numeric studies were performed in full detail.

Let us quote some concluding remarks:

1. In the context of the SM or its extensions without right-handed currents, the quark mass matrices 

$M_u$ and $M_d$ can be taken to be Hermitian without loss of generality. Non Hermitian quark mass matrices are relevant only 

when physics beyond the SM is being considered.

2. By counting free parameters we have that three texture zeros or less in the hermitian quark mass matrices of the SM, do 

not imply physical predictions for the elements of the flavor mixing matrix, because these texture zeros can always be obtained 

in a trivial way by using weak basis transformations.

3. Each one of the five RRR patterns studied here includes two physical relationships between the quark masses and the mixing 

angles, as can be seen from $V_{CKM}^\#;\;\;\; \#=I,II,\dots, V$. These predictions are case dependent, but far from being trivial.

4. Four and six texture zeros imply one and three physical relationships respectively, and allow us to write one or the 

three mixing angles as functions of the six quarks masses and the CP violation phase. But six texture zeros are already 

ruled out by analytic and numerical considerations~\cite{ripo, numal}.

5. More than six texture zeros are not possible because they imply either Det$M_q=0$ or a degenerate quark mass spectrum, 

both situations incompatible with the real world.

6. For the five patterns studied here, number IV (or equivalent number VII) reproduce all the experimental measured 

values, including the CP violation asymmetries, at $1\sigma$. For the other patterns, the $\beta$ angle appears 

at $2\sigma$. So, five texture zeros are far from being ruled out.

7. In our analysis for pattern IV, the several constraints for the strange quark mass imply  $0.051\leq m_s\mbox{(GeV)} \leq 0.056$.

\begin{center}

{\bf ACKNOWLEDGMENTS}

\end{center}

W.A.P thanks partial financial support from ``sostenibi\-lidad UdeA, 2013-14'', and R.H.B thanks partial finantial support from the ITM Research Center.

\appendix

\section{Mass values}

\label{aA}

For the experimental mass values used for carrying out the numerical calculations in the main text, we have adopted the 

following ranges of quark masses~\cite{Qexp} at the $M_Z$ energy scale (where the $V_{CKM}$ matrix elements are measured).

\vspace{1cm}

\begin{tabular}{||l|l||}\hline

Up sector & Down sector \\ \hline

$m_t=171.7\pm 3.0$ GeV & $m_b=2.89\pm 0.009$ GeV\\

$m_c=0.619\pm 0.084$ GeV & $m_s=55^{+16}_{-15}$ MeV \\

$m_u=1.27^{+0.5}_{-0.42}$ MeV & $m_d=2.90^{+1.24}_{-1.19}$ MeV \\ \hline\hline

\end{tabular}

\vspace{0.9cm}

The light quark masses $m_u,\; m_d$ and $m_s$ can be further constrained using the mass ratios~\cite{leut}

\begin{equation}\label{mara}
m_u/m_d=0.553\pm 0.043  \hspace{0.5cm} m_s/m_d=18.9\pm 0.8\;.
\end{equation}

Notice also that due to the experimental errors

\[m_t\pm m_c\pm m_u\approx m_t.\]

\section{Orthogonal Matrices}

\label{bB}

In this appendix we derive the analytic orthogonal matrices which diagonalize the several $3\times 3$ real quark mass matrices coming from the five different RRR patterns. As can be seen, the ten quark mass  matrices in Table~(\ref{T1}) can be grouped in only five different forms.

For the analysis we write the parameters of the several real mass matrices as functions of the quark mass eigenvalues $m_1,\;m_2$ and $m_3$, with the hierarchy $|m_1|<|m_2|<|m_3|$, by making use of the following invariant forms: tr$[M_q]$, tr$[(M_q)^2]$, and det$[M_q]$.

\subsection{Bi-diagonal form}

For this form the hermitian quark mass matrix is

\begin{equation}\label{ddf}
M_q^{dd} =\left(\begin{array}{ccc} 

0 & 0 & |a_q|e^{i\alpha_q^d} \\ 

0 & c_q & 0  \\

|a_q|e^{-i\alpha_q^d} & 0  & d_q

\end{array}\right),
\end{equation}

where the non-zero elements are only in the two diagonals. From the Table, only matrix $M_u^{V}$ belongs to this form.

Since det$[M_q^{dd}]=-c_q|a_q|^2<0$ for $c_q>0$, we use for the diagonal quark mass matrix Dg$(m_1,m_2,-m_3)$. 

Now using $U_q^{dd}=Dg.(1,1,e^{i\alpha_q^d})$ we have

\begin{equation}\label{rdd}
U_{q}^{dd}M_q^{dd}[U_q^{dd}]^{-1}=\left(\begin{array}{ccc} 

0 & 0& |a_q|  \\

0 & c_q & 0  \\

|a_q| & 0 & d_q
\end{array}\right)=M^{dd}_{q(R)}.
\end{equation}

The 3 matrix invariants provide us with the following set of equations

\begin{equation}\label{invdd}
\begin{split}
c_q+d_q&=m_1-m_2+m_3,\\
|a_q|^2c_q&=m_1m_2m_3,\\
|a_q|^2-c_qd_q&=m_1m_2+m_2m_3-m_1m_3,
\end{split}
\end{equation}
which allow us to write

\begin{equation}\label{idd}
d_{q}=m_1-m_3,\;\;c_q=m_2,\;\;|a_q|=\sqrt{m_1 m_3}
\end{equation}

with the exact diagonalizing transformation $O_{q}^{dd1}$ given by

\begin{equation}\label{odd}
O_{q}^{dd}=\left(
\begin{array}{ccc}

 \frac{\sqrt{m_3}}{\sqrt{m_3+m_1}} & 0 & \pm\frac{\sqrt{m_1}}{\sqrt{m_1+m_3}} \\

0 & 1 & 0 \\

\mp\frac{\sqrt{m_1}}{\sqrt{m_1+m_3}} & 0 & \frac{\sqrt{m_3}}{\sqrt{m_3+m_1}}

\end{array}
\right),
\end{equation}

which implies a mixing between the first and third quark families proportional to

\[\sin\theta_{13}=\sqrt{\frac{m_1}{m_1+m_3}}.\]

\subsection{Block Form}

For this form, the hermitian quark mass matrix is:

\begin{equation}\label{bf}
M_q^{bf} =\left(\begin{array}{ccc} 

0 & |a_q|e^{i\alpha_q^b} & 0 \\ 

|a_q|e^{-i\alpha_q^b} & c_q & 0  \\

0 & 0  & d_q
\end{array}\right).
\end{equation}

From the Table, matrices $M_d^{I},\; M_u^{II}$ and $M_d^{IV}$ belong to this form.

Using $U_q^{bf}=Dg.(1,e^{i\alpha_q^b},1)$ we have

\begin{equation}\label{bfr}
U_{q}^{bf}M_q^{bf}[U_q^{bf}]^{-1}=\left(\begin{array}{ccc} 

0 & |a_q| & 0 \\

|a_q| & c_q & 0  \\

0 & 0 & d_q
\end{array}\right)=M^{bf}_{q(R)}.
\end{equation}

We use for the diagonal quark mass matrix Dg$(m_1,-m_2,m_3)$. The $3\times 3$ matrix invariants combined with the hierarchy $|m_1|<|m_2|<|m_3|$ allow us to write

\begin{equation}\label{ibf}
\begin{split}
d_q&=m_3, \\ 
c_q&=m_1-m_2,\\ 
|a_q|&=\sqrt{m_1m_2},
\end{split}
\end{equation}

with the solution associated with the exact diagonalizing transformation $O_q^{(bf)}$ given now by

\begin{equation}\label{obf}
O_q^{bf}= \left(\begin{array}{ccc}

     \sqrt{\frac{m_2}{m_1+m_2}} & \pm\sqrt{\frac{m_1}{m_1+m_2}}  & 0 \\

   \mp\sqrt{\frac{m_1}{m_1+m_2}}   &  \sqrt{\frac{m_2}{m_1+m_2}}   & 0 \\

        0 & 0 & 1 
       \end{array}\right).
\end{equation}

\subsection{Nearest Neighbor form}

For this form, the hermitian quark mass matrix is:

\begin{equation}\label{nnf}
M_q^{nn} =\left(\begin{array}{ccc} 

0 & |a_q|e^{i\alpha_q^n} & 0 \\ 

|a_q|e^{-i\alpha_q^n} & 0 & |b_q|e^{i\beta_q^n}  \\

0 & |b_q|e^{-i\beta_q^n}  & d_q

\end{array}\right).
\end{equation}

From the Table, only $M_u^{III}$ belong to this form.

Using $U_q^{nn}=Dg.(1,e^{i\alpha_q^n},e^{i(\alpha_q^n+\beta_q^n)})$ we have

\begin{equation}\label{nnr}
U_{q}^{nn}M_q^{nn}[U_q^{nn}]^{-1}=\left(\begin{array}{ccc} 

0 & |a_q| & 0 \\ 

|a_q| & 0 & |b_q|  \\

0 & |b_q| & d_q
\end{array}\right)=M^{nn}_{q(R)}.\end{equation}

Using the definitions

\begin{equation}\label{mdiag}
M_{q(nn)}^{{\rm diag}}=[O_q^{nn}]^TM_{q(R)}^{nn} O_q^{nn}={\rm Diag}(m_1,-m_2,m_3),
\end{equation}

where the sub-indices 1,2,3 in the diagonal forms refer respectively to the masses for the quarks $u,c$ and $t$ for the up sector, as well as $d,s$ and $b$ for the down sector.

The three matrix invariants allow us to write now:

\begin{equation}\label{nnfi}
\begin{split}
d_q&=m_1-m_2+m_3,\\ 
|a_q|^2&=\frac{m_1m_2m_3}{m_1-m_2+m_3},\\ 
|b_q|^2&=\frac{(m_3-m_2)(m_3+m_1)(m_2-m_1)}{m_1-m_2+m_3},
\end{split}
\end{equation}

and the exact diagonalizing transformation $O_q^{nn}$ for this particular form is expressed as

\begin{widetext}

{\small

\begin{equation}\label{ronf}
O_q^{nn}=\left(\begin{array}{ccc} \pm\sqrt{\frac{m_2m_3(m_3-m_2)}{(m_3-m_1)(m_2+m_1)(m_1-m_2+m_3)}} & \pm\sqrt{\frac{m_1(m_3-m_2)}{(m_3-m_1)(m_1+m_2)}} &  \mp\sqrt{\frac{m_1(m_2-m_1)(m_1+m_3)}{(m_3-m_1)(m_1+m_2)(m_1-m_2+m_3)}} \\ 

\pm\sqrt{\frac{m_1m_3(m_1+m_3)}{(m_2+m_1)(m_3+m_2)(m_1-m_2+m_3)}}  & \mp\sqrt{\frac{m_2(m_1+m_3)}{(m_2+m_3)(m_1+m_2)}} & \pm\sqrt{\frac{m_2(m_3-m_2)(m_2-m_1)}{(m_2+m_1)(m_3+m_2)(m_1-m_2+m_3)}}\\ 

\pm\sqrt{\frac{m_1m_2(m_2-m_1)}{(m_2+m_3)(m_3-m_1)(m_1-m_2+m_3)}} &

\pm\sqrt{\frac{m_3(m_2-m_1)}{(m_2+m_3)(m_3-m_1)}} & 

\pm\sqrt{\frac{m_3(m_3-m_2)(m_1+m_3)}{(m_2+m_3)(m_3-m_1)(m_1-m_2+m_3)}}\end{array}\right),
\end{equation}}

\end{widetext}

where one has the freedom to choose two equivalent possibilities of phases (the up or down signs).

For the up quark sector, and due to the fact that $m_t\gg m_c\gg m_u$, the former matrix (\ref{ronf}) can be expanded as

\begin{widetext}

{\small

\begin{equation}\label{ronex}
O_u^{nn}\approx\left(\begin{array}{ccc} 

\pm (1-m_{uc}/2) & \pm\sqrt{m_{uc}}(1-m_{ct}/2-m_{uc}/2) & \mp\sqrt{m_{ut}}(1-m_{uc}+m_{ct}/2)  \\ 

\pm\sqrt{m_{uc}}(1-m_{uc}/2) & \mp(1-m_{ct}/2-m_{uc}/2) & \pm\sqrt{m_{ct}}(1-m_{uc}-m_{ct}/2)\\ 

\pm m_{ct}\sqrt{m_{ut}}  & \pm\sqrt{m_{ct}}(1-m_{uc}/2-m_{ct}/2) & \pm(1-m_{ct}/2) \end{array}\right),
\end{equation}}
\end{widetext}

\noindent

where $m_{ij}\equiv m_i/m_j,\;\; i<j;\;\; i,j=1,2,3=u,c,t$ respectively, and we have taken $m_{ut}=0$.

\subsection{Weak basis form}

For this form, the hermitian quark mass matrix is:

\begin{equation}\label{wbf}
M_q^{wb} =\left(\begin{array}{ccc} 

0 & |a_q|e^{i\alpha_q^w} & 0 \\ 

|a_q|e^{-i\alpha_q^w} & c_q & |b_q|e^{i\beta_q^w}  \\

0 & |b_q|e^{-i\beta_q^w}  & d_q

\end{array}\right).
\end{equation}

From the Table, matrices $M_u^{I},\; M_d^{II},\; M_d^{III}$ and $M_d^{V}$ belong to this form.

Using $U_q^{wb}=Dg.(1,e^{i\alpha_q^w},e^{i(\alpha_q^w+\beta_q^w)})$ we have

\begin{equation}\label{wbr}
U_{q}^{wb}M_q^{wb}[U_q^{wb}]^{-1}=\left(\begin{array}{ccc} 

0 & |a_q| & 0 \\

|a_q| & c_q & |b_q|  \\

0 & |b_q| & d_q

\end{array}\right)=M^{wb}_{q(R)}.
\end{equation}

The matrix invariants allow now to write:

\begin{equation}\label{cq4}
\begin{split}
c_q&=m_1-m_2+m_3-d_q,\\ 
a_q^2&=\frac{m_1m_2m_3}{d_q},\\ 
b_q^2&=\frac{(d_q-m_1)(d_q+m_2)(m_3-d_q)}{d_q},
\end{split}
\end{equation}

where, we use for the diagonal quark mass matrix Dg$(m_1,-m_2,m_3)$. Notice here that, for $d_q=m_3$, $b_q=0$ 

and thus $M_q^{wb}$ acquires the block form with an extra texture zero.

The exact diagonalizing transformation $O_q^{wb}$ can be expressed now as~\cite{verma}

{\small

\begin{widetext}

\begin{equation}\label{rotwb}
O_q^{wb}=\left(\begin{array}{ccc} 

\pm\sqrt{\frac{m_2m_3(d_q-m_1)}{(m_3-m_1)(m_2+m_1)d_q}} & \pm\sqrt{\frac{m_1m_3(d_q+m_2)}{(m_2+m_1)(m_3+m_2)d_q}} & \pm\sqrt{\frac{m_1m_2(m_3-d_q)}{(m_2+m_3)(m_3-m_1)d_q}} \\ 

\pm\sqrt{\frac{m_1(d_q-m_1)}{(m_3-m_1)(m_1+m_2)}} & \mp\sqrt{\frac{m_2(d_q+m_2)}{(m_2+m_3)(m_1+m_2)}} & \pm\sqrt{\frac{m_3(m_3-d_q)}{(m_2+m_3)(m_3-m_1)}}\\ 

\mp\sqrt{\frac{m_1(m_3-d_q)(m_2+d_q)}{(m_3-m_1)(m_1+m_2)d_q}} &

\pm\sqrt{\frac{m_2(m_3-d_q)(d_q-m_1)}{(m_2+m_1)(m_3+m_2)d_q}} & 

\pm\sqrt{\frac{m_3(d_q-m_1)(d_q+m_2)}{(m_2+m_3)(m_3-m_1)d_q}}\end{array}\right)^T.
\end{equation}
\end{widetext}}

\subsection{See-saw form}

For this form, the hermitian quark mass matrix is:

\begin{equation}\label{fss}
M_q^{ss} =\left(\begin{array}{ccc} 

0 & 0 & |a_q|e^{i\alpha_q^s} \\ 

0 & c_q & |b_q|e^{i\beta_q^s}  \\

|a_q|e^{-i\alpha_q^s} & |b_q|e^{-i\beta_q^s}  & d_q
\end{array}\right).
\end{equation}

From the Table, only matrix $M_u^{V}$ belongs to this form. Now, for the diagonal quark mass matrix we use: Dg$(-m_1,m_2,m_3)$.

Using now $U_q^{ss}=Dg.(1,e^{i(\alpha_q^s-\beta_q^s)},e^{i\alpha_q^s})$ we have

\begin{equation}\label{ssr}
U_{q}^{ss}M_q^{ss}[U_q^{ss}]^{-1}=\left(\begin{array}{ccc} 

0 & 0& |a_q|  \\

0 & c_q & |b_q|  \\

|a_q| & |b_q| & d_q
\end{array}\right)=M^{ss}_{q(R)}.
\end{equation}

The $3\times 3$ matrix invariants allow us to write

\begin{equation}\label{iss}
\begin{gathered}
|b_{q}|^2=\frac{(m_3+m_2-d_q)(d_q-m_2+m_1)(-d_{q}+m_3-m_1)}{-d_{q}-m_1+m_2+m_3},\\
c_{q}=-d_{q}-m_1+m_2+m_3,\\
|a_{q}|^2=\frac{m_1 m_2 m_3}{-d_{q}-m_1+m_2+m_3},
\end{gathered}
\end{equation}

where notice that for $d_q=m_3-m_1$, $b_q=0$ and $M_q^{ss}$ becomes the two diagonal form with an extra texture zero.

The exact diagonalizing transformation $O_{q}^{ss}$ is given now by

{\small

\begin{widetext}

\begin{equation}\label{oss}
O_{q}^{ss}=\left(\begin{array}{ccc}

-\sqrt{\frac{m_2 m_3\left(-d_{q}+m_2+m_3\right)}{\left(m_1+m_2\right) \left(m_1+m_3\right) \left(-d_{q}-m_1+m_2+m_3\right)}} & 

-\sqrt{\frac{m_1 \left(d_{q}-m_2+m_1\right)

\left(-d_{q}+m_3+m_1\right)}{\left(m_1+m_2\right) \left(m_1+m_3\right) \left(-d_{q}-m_1+m_2+m_3\right)}} & 

\sqrt{\frac{m_1 \left(-d_{q}+m_2+m_3\right)}{\left(m_1+m_2\right)

\left(m_1+m_3\right)}} \\

\sqrt{\frac{m_1 m_3 \left(-d_{q}+m_3-m_1\right)}{\left(m_1+m_2\right) \left(-d_{q}-m_1+m_2+m_3\right) \left(-m_2+m_3\right)}} &

 \sqrt{\frac{m_2 \left(d_{q}+m_1-m_2\right)

\left(-d_{q}+m_2+m_3\right)}{\left(m_1+m_2\right) \left(-d_{q}-m_1+m_2+m_3\right) \left(-m_2+m_3\right)}} & 

\sqrt{\frac{m_2 \left(-d_{q}-m_1+m_3\right)}{\left(m_1+m_2\right)

\left(-m_2+m_3\right)}} \\

\sqrt{\frac{m_1 m_2 \left(d_{q}-m_2+m_1\right)}{\left(-m_2+m_3\right) \left(-d_{q}-m_1+m_2+m_3\right) \left(m_1+m_3\right)}} &

 \sqrt{\frac{m_3\left(-d_{q}+m_2+m_3\right)

\left(-d_{q}-m_1+m_3\right)}{\left(-d_{q}-m_1+m_2+m_3\right) \left(-m_2+m_3\right) \left(m_1+m_3\right)}} & 

\sqrt{\frac{\left(d_{q}-m_2+m_1\right) m_3}{\left(-m_2+m_3\right)

\left(m_1+m_3\right)}}
\end{array}
\right).
\end{equation}

\end{widetext}}

\pagebreak

\end{document}